\documentclass{article}
\usepackage[utf8]{inputenc} 
\usepackage{kpfonts}
\usepackage{graphicx} 
\usepackage[margin=1.00in]{geometry} 
\usepackage{url}
\usepackage{xcolor}
\usepackage{tabularx}
\usepackage{multirow}
\usepackage{amsfonts}
\usepackage{amssymb}
\usepackage{pifont}
\newcommand{\cmark}{\ding{51}}%
\newcommand{\xmark}{\ding{55}}%
\usepackage{hyperref} 
\usepackage{marginnote}
\usepackage{authblk}
\usepackage{mathtools}
\let\svthefootnote\thefootnote
\newcommand\freefootnote[1]{%
  \let\thefootnote\relax%
  \footnotetext{#1}%
  \let\thefootnote\svthefootnote%
}
\usepackage[htt]{hyphenat}
\usepackage{subcaption}
\usepackage{adjustbox}

\title{PyTSC: A Unified Platform for Multi-Agent Reinforcement Learning in Traffic Signal Control}
\author[1]{Rohit Bokade}
\author[1]{Xiaoning Jin}
\affil[1]{Department of Mechanical and Industrial Engineering, Northeastern University, Boston, MA 02115, USA}
\date{\today}

\begin{document}

\maketitle
\begin{abstract}

Multi-Agent Reinforcement Learning (MARL) presents a promising approach for addressing the complexity of Traffic Signal Control (TSC) in urban environments. However, existing platforms for MARL-based TSC research face challenges such as slow simulation speeds and convoluted, difficult-to-maintain codebases. To address these limitations, we introduce PyTSC, a robust and flexible simulation environment that facilitates the training and evaluation of MARL algorithms for TSC. PyTSC integrates multiple simulators, such as SUMO and CityFlow, and offers a streamlined API, empowering researchers to explore a broad spectrum of MARL approaches efficiently. PyTSC accelerates experimentation and provides new opportunities for advancing intelligent traffic management systems in real-world applications.

\end{abstract}

\section{Introduction}

Effective Traffic Signal Control (TSC) is fundamental to urban traffic management, responsible for guiding the movement of vehicles through intersections by controlling traffic lights. The primary goals of TSC are to minimize traffic congestion, enhance traffic flow, and improve safety for both vehicles and pedestrians. Poor TSC optimization leads to increased congestion, fuel consumption, and pollution. Longer wait times at signals lead to increased fuel consumption, which not only exacerbates environmental issues through higher emissions but also results in economic losses due to delays. Moreover, inefficient TSC negatively impacts the quality of life in urban areas, contributing to increased noise and air pollution.

Multi-Agent Reinforcement Learning (MARL) offers a promising approach to tackling these TSC challenges by allowing multiple agents to collaborate within a shared environment. In fully cooperative settings, agents work toward a common goal by interacting with their environment and with one another and refine their actions based on the feedback from the environment. MARL's versatility is demonstrated by its successful application in various domains \cite{du2021survey}. For instance, in robotics, MARL has been used to coordinate multiple robots in tasks such as search and rescue operations \cite{nguyen2020deep, canese2021multi}. These successes highlight MARL's potential to solve complex, multi-faceted problems, making it an ideal candidate for optimizing TSC in dynamic and unpredictable urban environments.

\subsection{Challenges in Current MARL Research for TSC}

The application of MARL to TSC has seen notable advancements \cite{noaeen2022reinforcement, chen2022real, wei2019survey, haydari2020deep}. Two simulators, SUMO \cite{SUMO2018} and CityFlow \cite{zhang2019cityflow}, are widely recognized in this domain, and several open-source tools have been developed to leverage these platforms \cite{sumorl, ault2021reinforcement, mei2022libsignal}. Recent efforts have aimed at merging the TSC environments of both simulators and unifying domain metrics, standardizing evaluation criteria and providing a consistent framework for problem formulation in TSC research \cite{mei2022libsignal}. Benchmarks tailored for specific datasets in SUMO have also been developed, supporting a diverse range of MARL algorithms \cite{ault2021reinforcement}.

Despite progress in applying Multi-Agent Reinforcement Learning (MARL) to Traffic Signal Control (TSC), the research community still lacks a unified, modular, and extensible platform that meets the needs of modern MARL algorithms. Most existing TSC simulators are tightly coupled to their frameworks and are not designed to support the flexible integration of advanced MARL methodologies, particularly frameworks like Centralized Training Decentralized Execution (CTDE), which have gained significant traction in recent years.

While simulators such as SUMO and CityFlow are widely used, their respective TSC libraries are neither optimized for CTDE-based architectures nor compatible with popular MARL libraries like EPyMARL \cite{papoudakis2020benchmarking} and MARLLib \cite{hu2022marllib}. This lack of compatibility limits experimentation and the exploration of powerful frameworks that can balance centralized learning with decentralized execution, which is crucial for efficient and scalable traffic management in real-world systems.

Furthermore, the absence of a streamlined and modular design in existing tools makes them challenging to extend and adapt for new research directions. Researchers often spend considerable time grappling with integration and code maintenance, rather than focusing on the development and testing of novel MARL algorithms. This inefficiency delays progress and discourages the widespread adoption of cutting-edge MARL techniques in TSC.

To address these gaps, we propose PyTSC, a library specifically designed to overcome these limitations. PyTSC offers a clean, modular, and extensible environment that seamlessly integrates with CTDE frameworks like EPyMARL and MARLLib, enabling researchers to quickly prototype, train, and evaluate MARL algorithms in a traffic control context. By providing a robust and flexible platform, PyTSC empowers the research community to explore novel TSC solutions, leading to faster experimentation, greater reproducibility, and ultimately, more effective traffic signal control strategies.

\subsection{Contributions}

This work introduces PyTSC, a flexible and efficient simulation environment designed to address these challenges in the MARL-TSC research domain. PyTSC fills a crucial gap in the TSC research ecosystem by providing a platform that simplifies integration, supports cross-simulator comparisons, and offers a clean interface for deploying advanced MARL methods like CTDE. The key contributions of this research are:
\begin{enumerate}
    \item \textbf{Compatibility with Multiple Simulators:} PyTSC supports both SUMO and CityFlow simulators, offering a consistent API. This design facilitates the integration of additional simulators in the future.
    \item \textbf{Optimized for Speed:} The \texttt{Retriever} module in PyTSC efficiently gathers required information from the simulator after each time step, minimizing simulator queries and enhancing simulation speeds\footnote{\url{https://sumo.dlr.de/docs/FAQ.html\#traci}}.
    \item \textbf{Leveraging Graphical MARL Techniques:} The \texttt{NetworkParser} module processes network files, extracting data crucial for graphical methods MARL algorithm development (e.g. adjacency matrix, centrality measures, etc.), facilitating the use of advanced graph neural networks in TSC.
    \item \textbf{Dataset Aggregation:} PyTSC aggregates commonly used datasets in MARL-TSC. The environment also features modules like \texttt{GridGenerator} and \texttt{TripGenerator} for synthetic network testing and trip generation, respectively.
    \item \textbf{Unified MARL Formulation for TSC:} While we recognize that we did not develop the Dec-POMDP and Networked MMDP formulations, we advocate for their use in TSC research. These formulations, which allow for decentralized control schemes, are comprehensive and well-suited for deep MARL techniques in TSC. By promoting these formulations, we aim to standardize nomenclature and problem formulations in the TSC field, facilitating clearer communication and collaboration among researchers.
    \item \textbf{Experiments with CTDE MARL Frameworks:} As demo, we present a experiments with several of state-of-the-art MARL techniques, which follow Centralized Training and Decentralized Execution (CTDE) paradigm, using the EPymarl library\cite{papoudakis2020benchmarking}.
\end{enumerate}
In summary, PyTSC is not merely a tool but a significant advancement in TSC research. By integrating MARL into a well-structured environment, PyTSC has the potential to redefine Traffic Signal Control research, propelling both academic inquiry and practical applications.

\section{The PyTSC Framework}

    \begin{figure}[!ht]
        \centering
        \includegraphics[width=\textwidth]{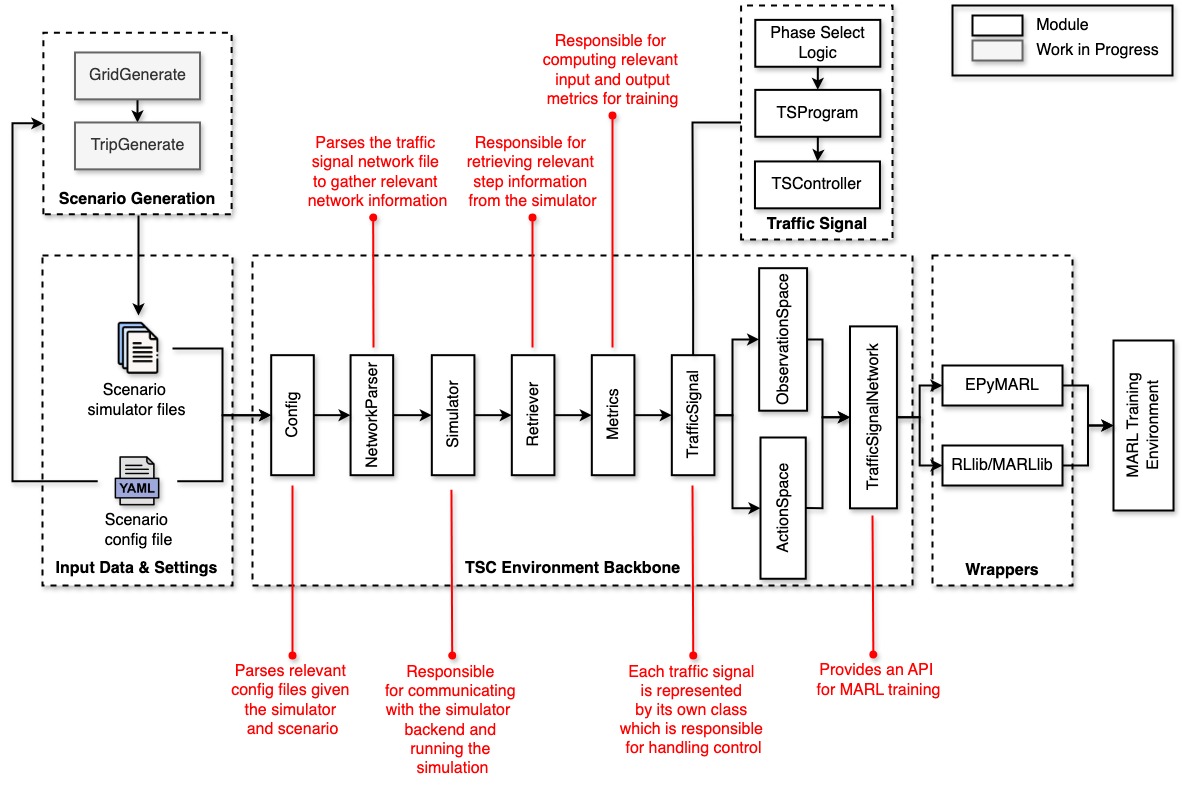}
        \caption{Overview of PyTSC Architecture}
        \label{fig:pytsc_architecture}
    \end{figure}

    The PyTSC Framework stands as a meticulously designed structure, bridging the gap between traffic simulations and Multi-Agent Reinforcement Learning (MARL). The library is available at \url{https://github.com/rbokade/pytsc}.This strategic design ensures that researchers can delve into algorithmic intricacies without the overhead of integration complexities. The framework's key attributes include:
    
    \paragraph{Support for Diverse Simulator Backends:} PyTSC integrates a consistent API for two renowned simulator backends: SUMO \cite{SUMO2018} and CityFlow \cite{zhang2019cityflow}. Simulator specific classes like \texttt{ConfigParser}, \texttt{Retriever}, \texttt{Simulator}, \texttt{TrafficSignal} allow for processing the input from the simulators into a common format which can then be used in the \texttt{TrafficSignalNetwork} environment class. This uniform interface not only maintains consistency across simulators but also offers a foundation for researchers to incorporate additional simulator backends as needed. 
    \paragraph{Customizable:} The modular framework of PyTSC serves as a comprehensive testbed for experimenting. Researchers can experiment with various traffic signal network settings by choosing from existing modules or extend them to suit their own needs. For example, the \texttt{TLSFreePhaseSelectLogic} and \texttt{TLSRoundRobinPhaseSelectLogic} allow users to select either adaptive phase selection or fix it to a round robin strategy. Users can also modify the information required by the MARL algorithm by simply extending \texttt{BaseObservationSpace}, \texttt{BaseActionSpace}, \texttt{BaseRewardFunction} according to their needs.
    \paragraph{Optimized Performance:} The framework is optimized to gather essential metrics from the backend simulator in a single query after each simulation step. This streamlined approach minimizes redundant queries, leading to faster simulation processes. For SUMO simulator backends, it uses subscriptions \url{https://sumo.dlr.de/docs/FAQ.html#traci} to further speed up simulations.
     \paragraph{Integration of Static Network Features:} Before simulation commencement, the \texttt{NetworkParser} module parses all associated network files of the chosen simulator. This provides researchers with added network insights, such as centrality metrics or adjacency matrices, which can be pivotal for enhancing MARL algorithm performance in Traffic Signal Control.
    \paragraph{Compatibility with MARL Training Frameworks:} PyTSC introduces a well-defined API under the module \texttt{TrafficSignalNetwork} that integrates seamlessly with established MARL libraries, such as rllib \cite{liang2018rllib} and pymarl \cite{samvelyan2019starcraft}, facilitating their application in TSC.

\section{Experiments}

\subsection{Traffic Signal Network Scenarios}

We have curated \textcolor{red}{10} open source scenarios most commonly used by researchers while applying MARL techniques to TSC. These scenarios, widely adopted by researchers in the field, encompass both synthetic and real-world traffic networks and are compatible with both CityFlow and SUMO simulators. A detailed overview of these scenarios is presented in Table \ref{tab:pytsc_scenarios}.
\freefootnote{\url{https://github.com/cts198859/deeprl\_network/}}
\freefootnote{\url{https://github.com/LucasAlegre/sumo-rl}}
\freefootnote{\url{https://github.com/traffic-signal-control/sample-code/}}
\freefootnote{\url{https://github.com/Pi-Star-Lab/RESCO/}}

\begin{figure}
    \centering
    \includegraphics[width=0.275\textwidth]{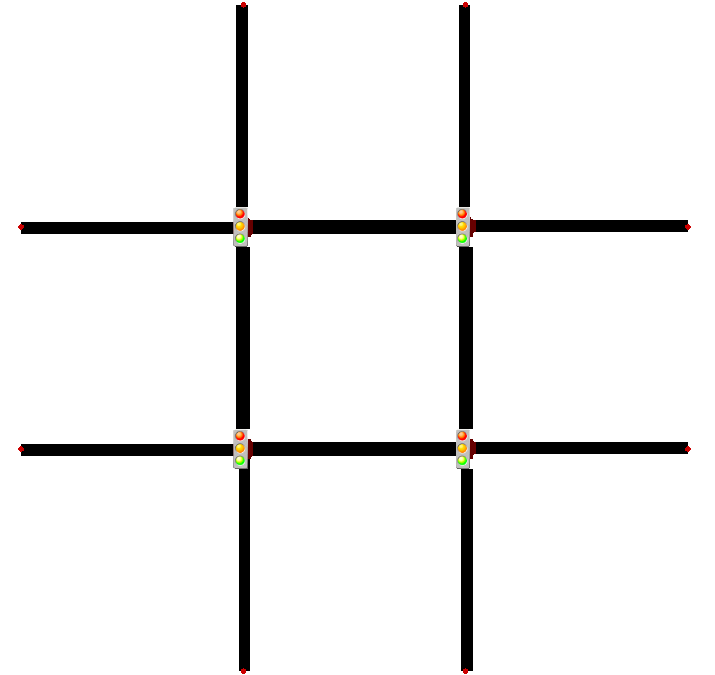}
    \includegraphics[width=0.265\textwidth]{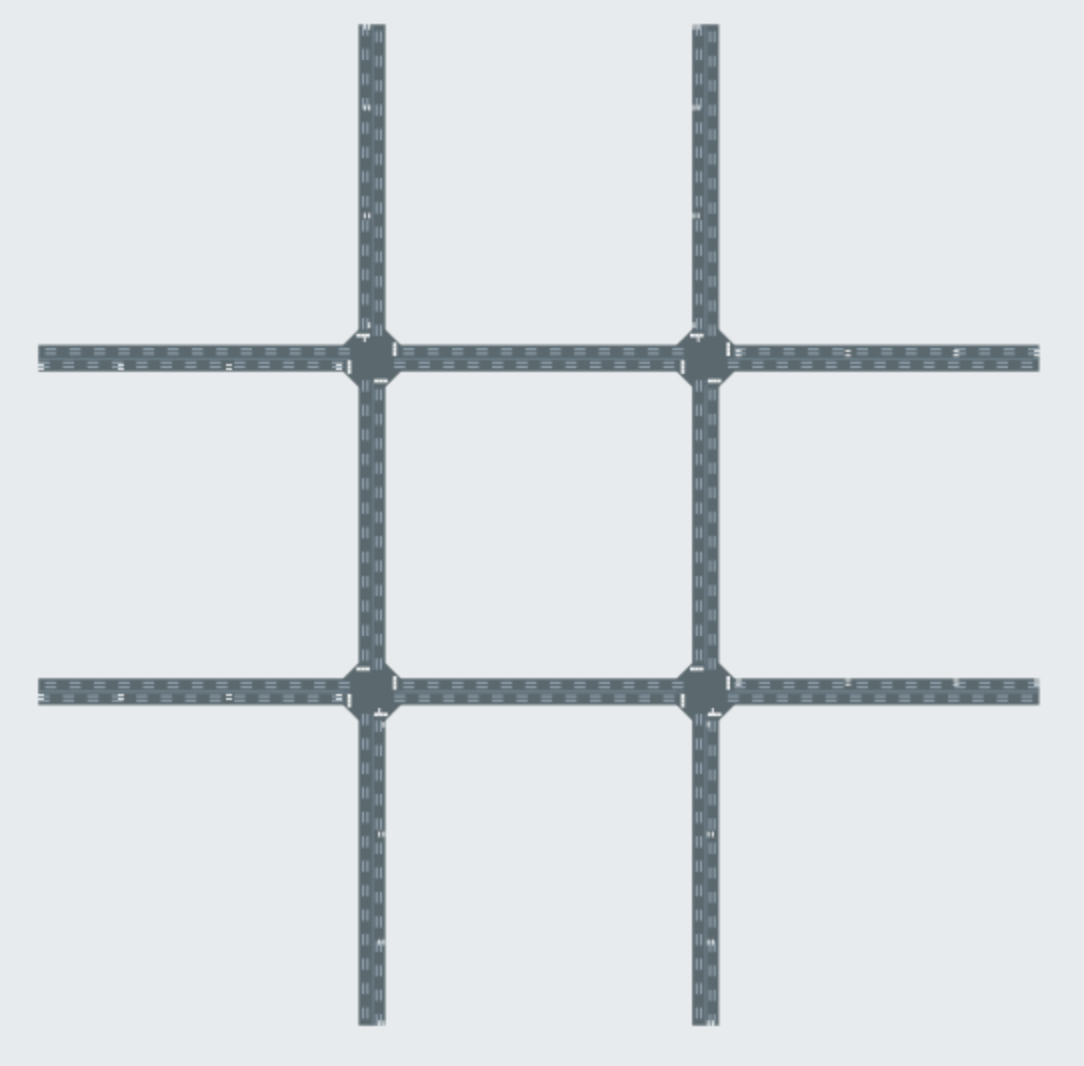}
    \label{fig:sa}
    \caption{$2\times2$ Grid SUMO (left) CityFlow (right)}
    \label{fig:enter-label}
\end{figure}

\begin{figure}
    \centering
    \includegraphics[width=0.270\textwidth]{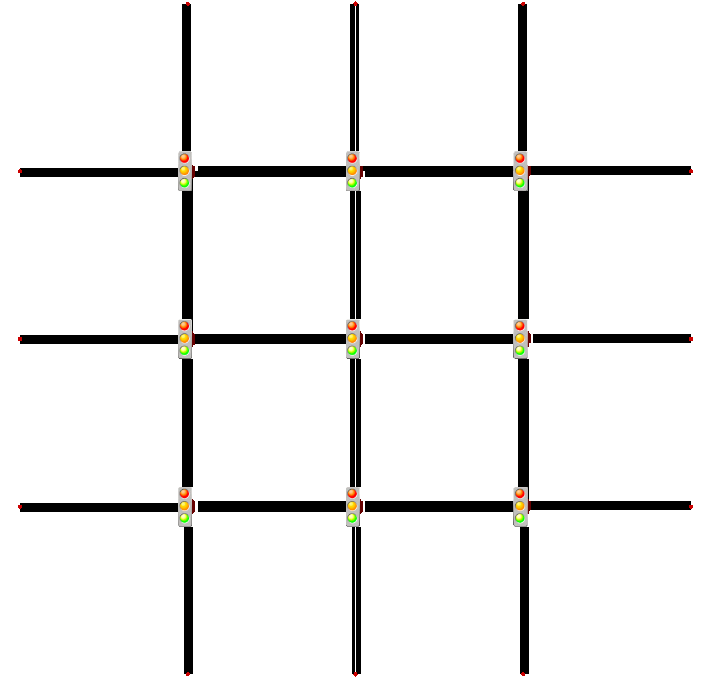}
    \includegraphics[width=0.265\textwidth]{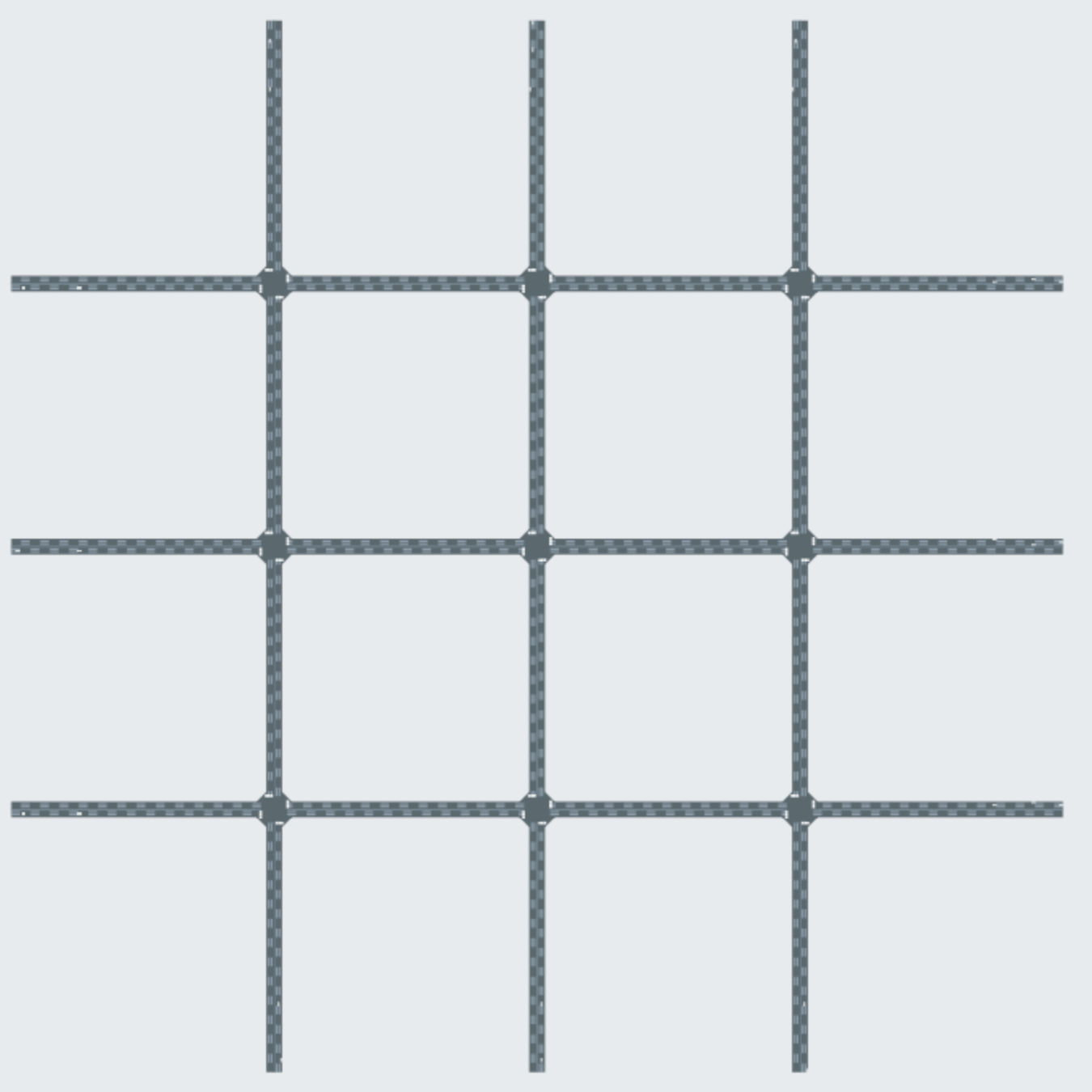}
    \label{fig:image_splitting_substrate}
    \caption{$3\times3$ Grid SUMO (left) CityFlow (right)}
\end{figure}

\begin{figure}[!ht]
    \centering
    \begin{subfigure}{0.35\textwidth}
        \includegraphics[width=\textwidth]{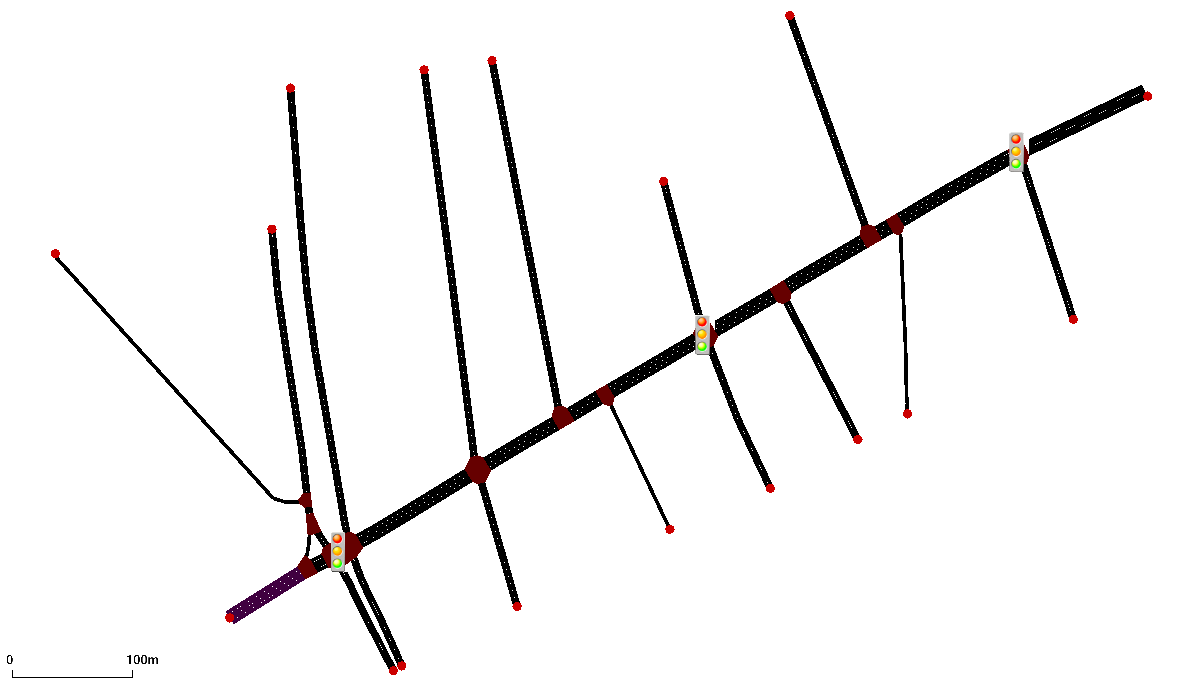}
        \label{fig:cologne3}
        \caption{Cologne (3 traffic signals)}
    \end{subfigure}
    \hspace{2em}
    \begin{subfigure}{0.20\textwidth}
        \includegraphics[width=\textwidth]{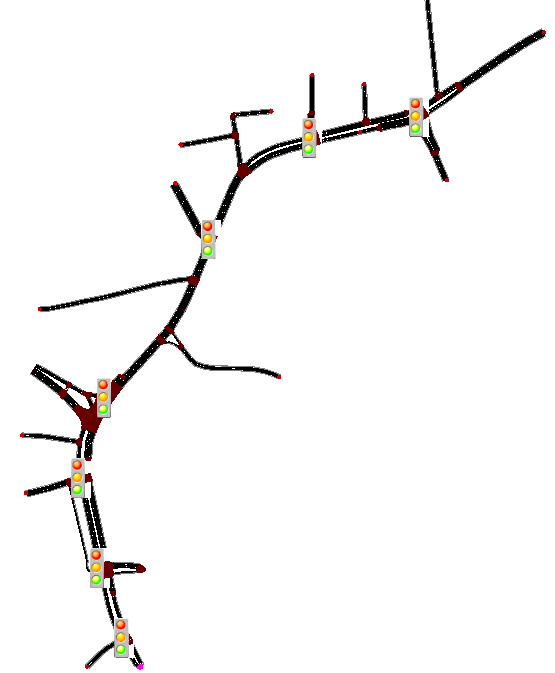}
        \label{fig:ingolstadt7}
        \caption{Ingolstadt}
    \end{subfigure}    
    \vspace{1em}
    \begin{subfigure}{0.45\textwidth}
        \includegraphics[width=\textwidth]{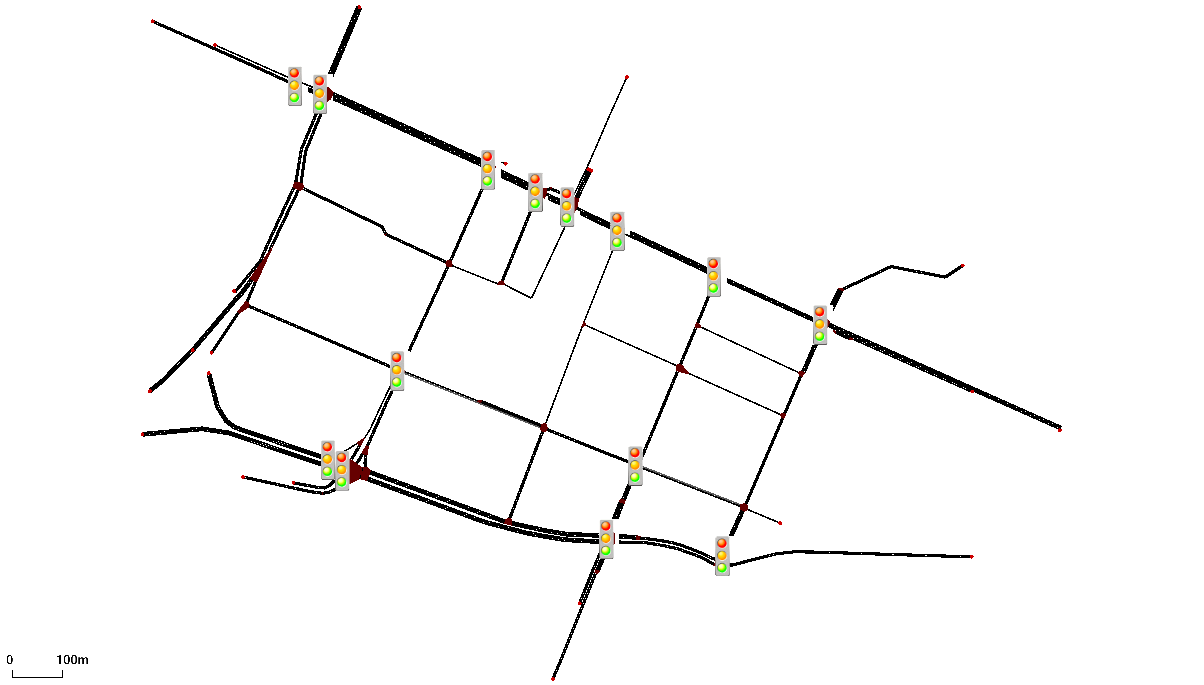}
        \caption{Pasubio}
        \label{fig:pasubio}
    \end{subfigure}
    \hspace{2em}
    \begin{subfigure}{0.25\textwidth}
        \includegraphics[width=\textwidth]{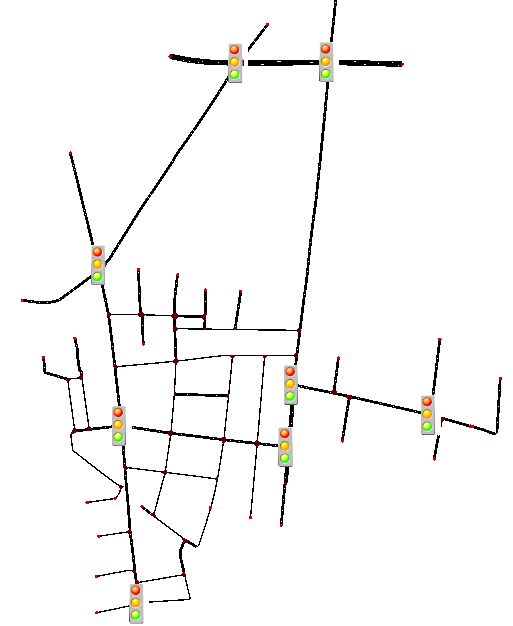}
        \caption{Cologne (8 traffic signals)}
        \label{fig:cologne8}
    \end{subfigure}
    \caption{Real-world environments for SUMO} 
    \label{fig:synthetic_envs}
\end{figure}

\begin{figure}[!ht]
    \centering
    \begin{subfigure}{0.275\textwidth}
        \includegraphics[width=\textwidth]{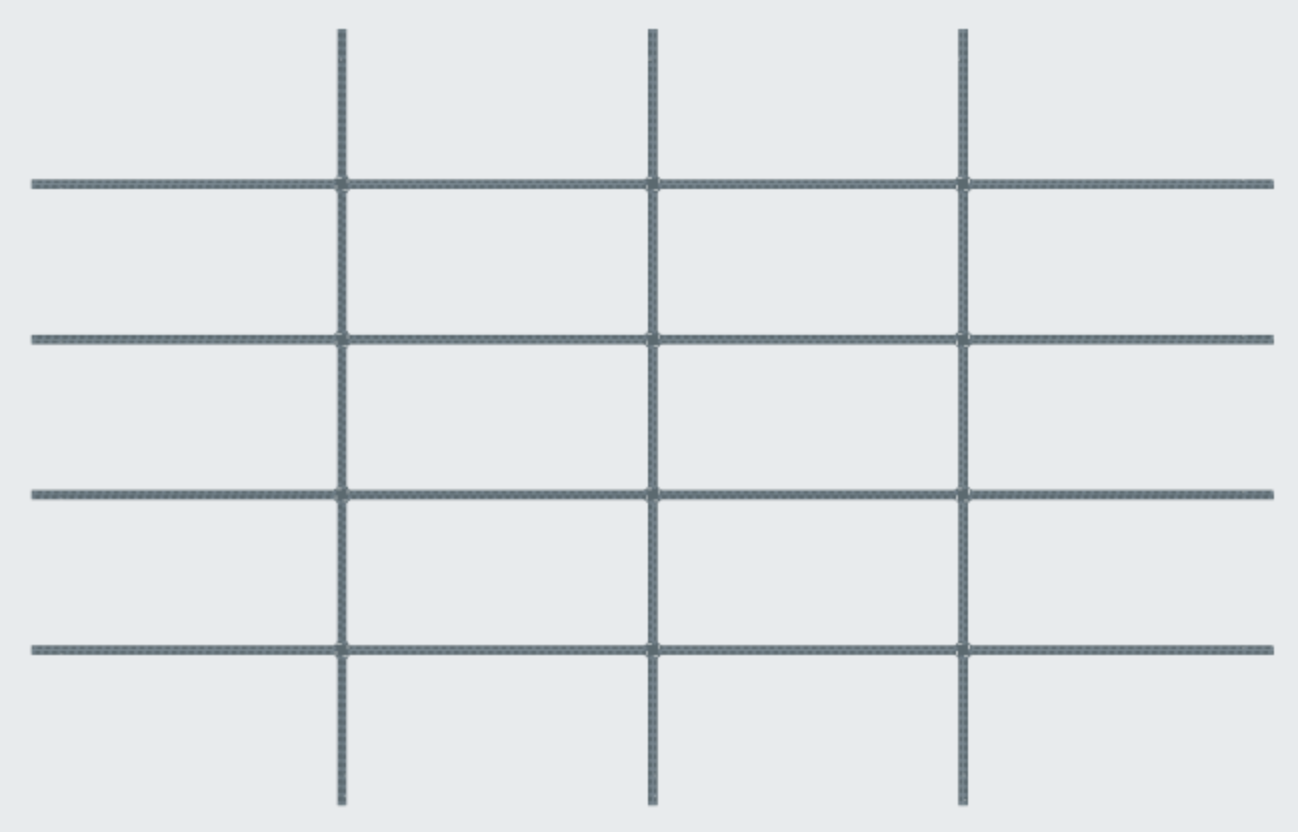}
        \label{fig:jinan}
        \caption{Jinan $3\times4$}
    \end{subfigure}
    \hspace{2em}
    \begin{subfigure}{0.275\textwidth}
        \includegraphics[width=\textwidth]{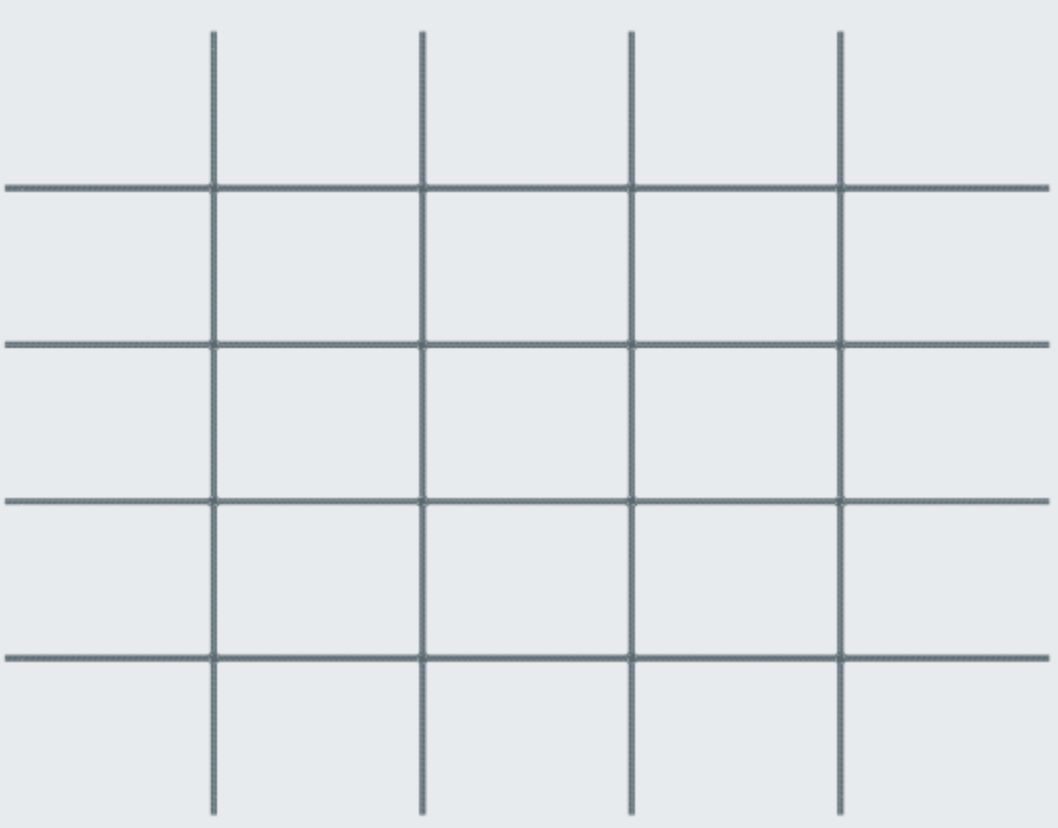}
        \caption{Hangzhou $4\times4$}
        \label{fig:hangzhou}
    \end{subfigure}
    \caption{Real-world environments for CityFlow} 
    \label{fig:cityflow_envs}
\end{figure}


\begin{table*}[!ht]
    \centering
    \begin{tabularx}{0.85\textwidth}{c c c c c}
        \hline
        \textbf{Simulator} & \textbf{Scenario} & \textbf{Network Type} & \textbf{Agent Type} & \textbf{Total agents} \\
        \hline 
        \multirow{7}{*}{\textbf{SUMO}} &
        $2 \times 2$ grid & Synthetic & Homogeneous & 4 \\ 
        & $3 \times 3$ grid & Synthetic & Homogeneous & 9 \\ 
        & Cologne 3 & Real-world & Heterogeneous & 3 \\ 
        & Ingolstadt 7 & Real-world & Heterogeneous & 7 \\ 
        & Cologne 8 & Real-world & Heterogeneous & 8 \\
        & Pasubio & Real-world & Heterogeneous & 8 \\
        \hline 
        \multirow{4}{*}{\textbf{CityFlow}} & 
        $2 \times 2$ grid & Synthetic & Homogeneous & 4 \\ 
        & $3 \times 3$ grid & Synthetic & Homogeneous & 9 \\ 
        & Jinan ($3 \times 4$ grid) & Real-world & Homogeneous & 12 \\ 
        & Hangzhou ($4 \times 4$ grid) & Real-world & Homogeneous & 16 \\
        \hline
    \end{tabularx}
    \caption{Scenarios included in PyTSC}
    \label{tab:pytsc_scenarios}
\end{table*}

The scenarios in Table \ref{tab:pytsc_scenarios} offer a mix of synthetic grid networks and real-world traffic settings. Synthetic grids, from $2 \times 2$ to $3 \times 3$, provide a controlled environment with homogeneous agents for testing MARL techniques. In contrast, real-world scenarios from cities like Cologne, Ingolstadt, and Monaco present urban complexities with heterogeneous agents, ranging from $3$ to $16$. This diversity ensures thorough testing of MARL across various scales. Additionally, the scenarios' compatibility with both SUMO and CityFlow allows researchers flexibility in simulation choices. Overall, these scenarios form a robust benchmarking platform for MARL in TSC, covering diverse network types and complexities.

\subsection{Traffic Signals as Agents}

In MARL under fully-cooperative settings, agents learn to achieve a common goal or maximize their individual rewards through interaction with the environment and each other. In the context of TSC, MARL provides a framework for developing traffic signal control strategies that can adapt to changing traffic patterns and optimize flow. Traffic signals are modeled as agents whose goal is to choose control the traffic lights to minimize congestion throughout the traffic signal network. 

\subsubsection{Decentralized Partially Observable Markov Decision Processes (Dec-POMDPs)}

Dec-POMDPs model multi-agent systems \cite{oliehoek2016concise} where agents interact in a decentralized way under limited visibility. This is pertinent for analyzing traffic signal control. \\ \\
\textbf{Definition} A Dec-POMDP is defined as a tuple 
$\langle N, S, \{A_i\}, \{O_i\}, T, \{\Omega_i\}, R \rangle$, where:
\begin{itemize}
    \item $N$: A finite set of \(n\) agents, $N \equiv \{1, \ldots, n\}$.
    \item $S$: A finite set of states that describe the global state of the environment.
    \item $\{A_{i}\}$: A finite set of actions for each agent $i$, with the joint action space $A \equiv \{A_{1} \times \ldots \times A_{n}\}$.
    \item $\{O_{i}\}$: A finite set of observations for each agent $i$, with the joint observation space $O \equiv \{O_{1} \times \ldots \times O_{n}\}$.
    \item $T: S \times A \mapsto \Delta(S)$: A state transition function that maps the current state and joint action to a probability distribution over next states.
    \item $\{\Omega_{i}: S \times A \mapsto \Delta(O_{i})\}$: An observation function for each agent, mapping the current state and joint action to a probability distribution over individual observations.
    \item $R: S \times A \mapsto \mathbb{R}$: A reward function that maps a global state and joint action to a real-valued reward.
\end{itemize}
By capturing the decentralized nature and partial observability inherent in urban traffic systems, Dec-POMDPs offer a foundation for designing MARL frameworks that can lead to more responsive and efficient traffic signal control. 

\subsubsection{Observation representation}

Each traffic signal has a limited range of vision of 50 meters, within which it can obtain information related to the traffic flow. This is equivalent to the sensory information that can be obtained from practical common sensors. The observation for each traffic signal consists of: the number of vehicles $\{ n_{l} \}_{l=1}^{L_{i}}$, the average normalized speed of the vehicles $\{ s_{l} \}_{l=1}^{L_{i}}$, the number of halted vehicles (queue lengths) $\{ q_{l} \}_{l=1}^{L_{i}}$, and the current \texttt{phaseID} of the traffic signal, where $L_{i} \in L$ are the incoming lanes for a traffic signal $i$ and $L$ is a set of all the lanes in the network.

\subsubsection{Action Representation}


For each traffic signal $i$, we define its action $a_{i}$ as choosing one green phase from a list of available phases. A traffic signal can select any green phase from its list or keep its current one, but it must then follow the next yellow phase, which is enforced by the environment. The action selection interval and the yellow phases are fixed for a duration of 5 simulation seconds.

\subsubsection{Reward}

Various metrics are used for rewards in traffic signal control settings. In our study, we chose queue length $q_{l}$ as the performance metric of the traffic signal controller due to its simplistic nature and its property of representing an instantaneous feedback signal. We define the objective function as minimizing the number of vehicles stopped throughout the network $$ $$ where $r_{t} \in \mathbb{R}$ is the global reward and $l \in L$ represents the lanes in the network.

\subsection{MARL Frameworks}

For benchmarking CTDE algorithms in TSC, we employ EPymarl, an extension of the widely recognized Pymarl library. EPymarl encompasses a broad spectrum of algorithms under the reinforcement learning paradigms of Q-learning, actor-critic, and policy gradient methods. Our selection focuses on the most prevalent MARL frameworks, as detailed in Table \ref{tab:marl_algorithms}.

\freefootnote{\url{https://github.com/uoe-agents/epymarl/}}
\freefootnote{\url{https://github.com/oxwhirl/pymarl}}

\begin{table*}[!ht]
    \centering
    \begin{tabularx}{0.9\textwidth}{c c c c c}
        \hline 
        \textbf{Algorithm} & \textbf{Centralized training} & \textbf{Off-/On-policy} & \textbf{Value-based} & \textbf{Policy-based} \\
        \hline 
        IQL & \xmark & \xmark & \cmark & \xmark \\ 
        IA2C & \xmark & \cmark & \cmark & \cmark \\ 
        VDN & \cmark & \xmark & \cmark & \xmark \\ 
        QMIX & \cmark & \xmark & \cmark & \xmark \\
        MAA2C & \cmark & \cmark & \cmark & \cmark \\
        \hline
    \end{tabularx}
    \caption{MARL Algorithms for Benchmarking}
    \label{tab:marl_algorithms}
\end{table*}

Analyzing the algorithms presented in Table \ref{tab:marl_algorithms}, we observe a diverse range of MARL techniques tailored for different problem settings. IQL, for instance, is a decentralized, value-based method that operates off-policy. In contrast, IA2C is both value and policy-based, functioning on-policy without centralized training. VDN and QMIX, while both being value-based, differ in their approach to centralized training, with QMIX employing it. MAA2C stands out as a versatile algorithm, incorporating both value and policy-based methods, operating on-policy, and utilizing centralized training. This selection ensures a comprehensive evaluation of MARL techniques across various training paradigms, policy orientations, and value determinations. By benchmarking these algorithms, we aim to provide insights into their applicability, strengths, and limitations within the context of TSC. 


\subsection{Evaluation Protocols}

The performance of the Multi-Agent Reinforcement Learning (MARL) algorithms was evaluated through a carefully designed training and testing process. Each episode during training was constrained to 360 simulation seconds, which corresponds to 72 time steps. This time frame was selected to ensure a balance between simulation length and computational efficiency. Therefore, one simulation hour consisted of 10 episodes, which allowed the MARL algorithms to interact with the environment multiple times within a relatively short period (see Table \ref{tab:benchmarking_metrics} for a detailed breakdown of time steps and simulation periods).

\begin{table*}[!ht]
    \centering
    \begin{tabularx}{0.75\textwidth}{c c c c}
        \hline 
        \textbf{Metric} & \textbf{Time step} & \textbf{Simulation seconds} & \textbf{Simulation hours} \\
        \hline 
        Step & 1 & 5 & 0.083 \\ 
        Episode limit & 72 & 360 & 0.10 \\ 
        Training (length) & 4.32M & 21.6M & 6000 \\ 
        Test (interval) & 14400 & 7200 & 2 \\
        Test (length) & 720 & 3600 & 1 \\
        \hline
    \end{tabularx}
    \caption{Breakdown of Hyperparameters Used for Benchmarking}
    \label{tab:benchmarking_metrics}
\end{table*}

In total, the algorithms were trained for 4.32 million time steps, which corresponds to 36,000 episodes. This extensive training schedule, equivalent to 6,000 hours of simulated traffic, provided ample opportunity for the agents to learn optimal strategies for traffic signal control. To assess the generalization capabilities of the trained models, evaluation was performed every 200 episodes. For each evaluation cycle, the models were tested over 10 episodes to ensure that performance was not merely due to overfitting but could be replicated under various conditions (see Table \ref{tab:benchmarking_metrics} for details on testing intervals and length).

To optimize the MARL algorithms, hyperparameter tuning was conducted on smaller synthetic grid networks, such as 2×2 and 3×3 grids. These simpler environments allowed for more efficient exploration of different hyperparameter combinations, ensuring that the best configurations were chosen before applying the algorithms to larger and more complex networks. The chosen hyperparameters for benchmarking are shown in Table \ref{tab:hyperparameters}.

\begin{table}[!ht]
    \centering
    \begin{tabular}{l c}
        \hline 
        \textbf{Hyperparameter} & \textbf{Value} \\
        \hline 
        Buffer size (episodes) & 5000 \\
        Hidden dimension & 64 \\
        Learning rate & 0.0005 \\
        Evaluation epsilon & 0.0 \\
        Epsilon anneal (steps) & 50000/100000 \\
        Target update (episodes) & 200 \\
        Entropy coeff. & 0.01 \\
        \hline
    \end{tabular}
    \caption{Hyperparameters}
    \label{tab:hyperparameters}
\end{table}

In addition to these learning metrics, several TSC-specific metrics were employed to capture the broader impact of the traffic signal control system. These included the total number of vehicles queued across the network, the average travel time, average occupancy, average speed, average delay, and average wait time.

All experiments were conducted on Northeastern University’s High-Performance Computing (HPC) Discovery cluster. Each experiment utilized 8 single-core CPUs, with 4 parallel environments running simultaneously to gather data efficiently. This setup allowed for significant computational power and parallelization, enabling the MARL algorithms to process multiple simulations concurrently and reduce the overall time required for training and evaluation.

\section{Results}


The performance of various MARL algorithms was evaluated across different environments using SUMO and CityFlow simulators. These environments ranged from simple synthetic grid networks to more complex real-world networks, such as Jinan, Hangzhou, Pasubio, and Cologne. The results highlight several key factors that affect algorithmic performance, including network topology and the simulation platform.

\subsection{Performance on Synthetic Grid Networks}

\begin{figure}[!ht]
    \centering
    \begin{subfigure}{0.375\textwidth}
        \includegraphics[width=\textwidth]{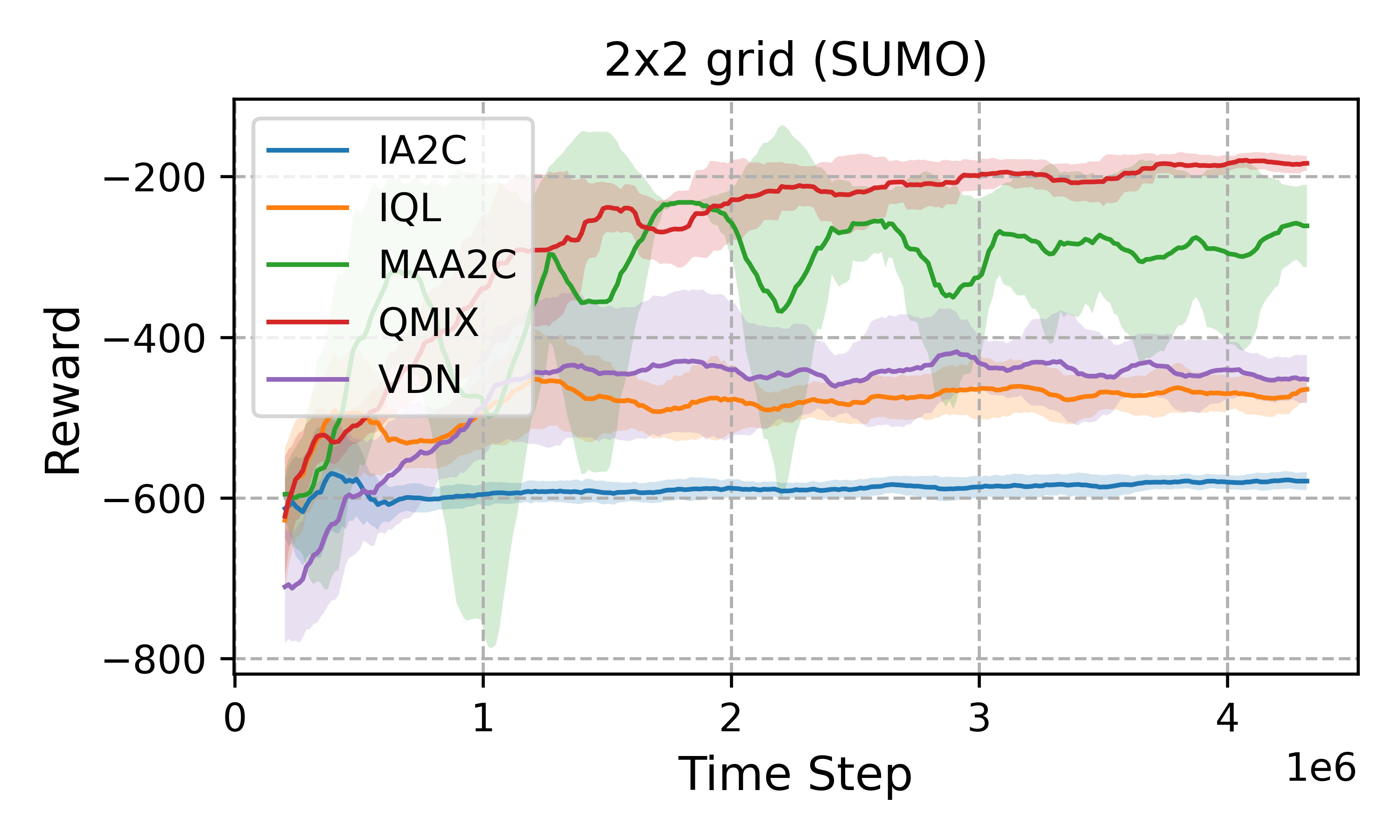}
        \label{fig:sample_thermal_image}
    \end{subfigure}
    \begin{subfigure}{0.375\textwidth}
        \includegraphics[width=\textwidth]{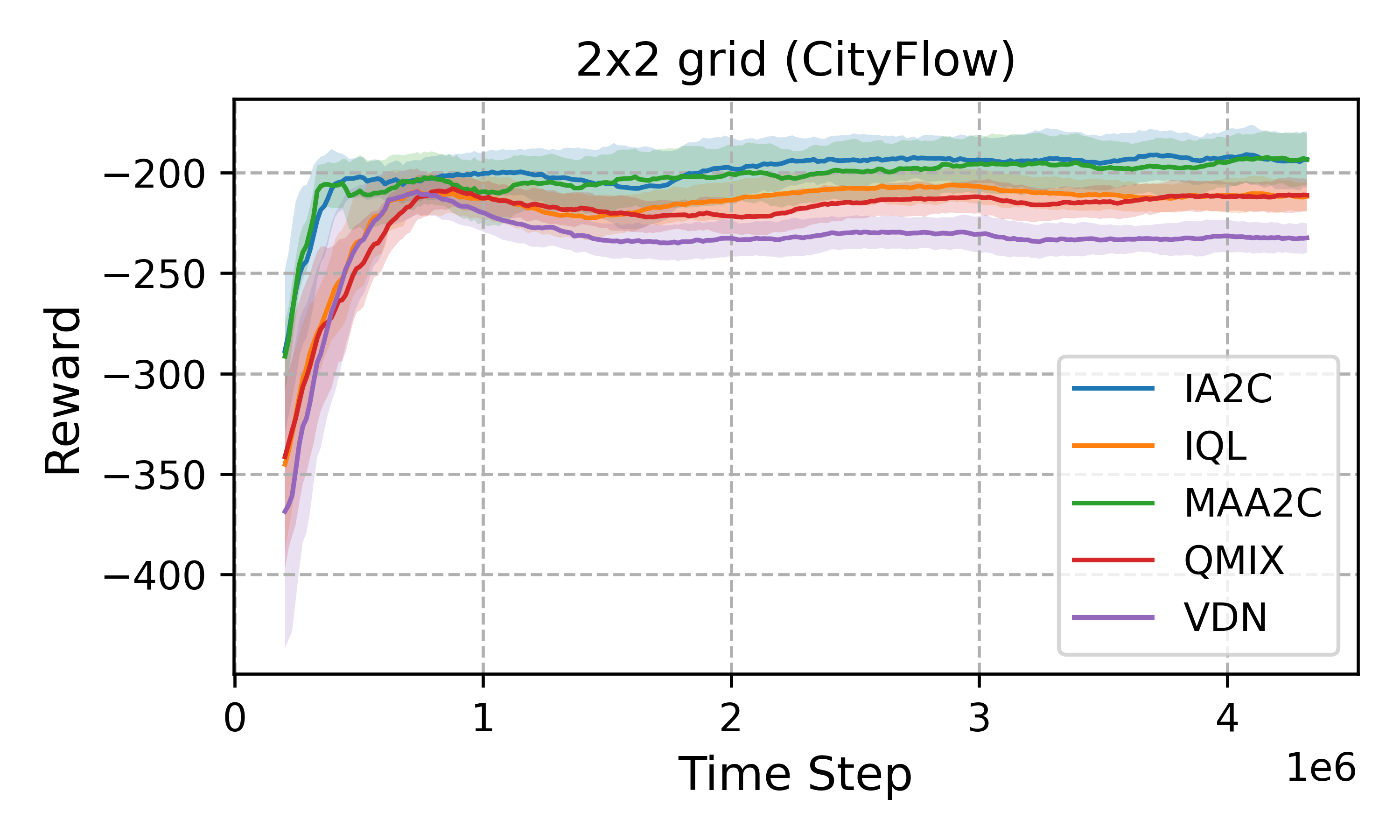}
        \label{fig:sample_thermal_image}
    \end{subfigure}
\end{figure}

In synthetic grid networks (2x2 and 3x3), centralized algorithms like QMIX and MAA2C outperform decentralized approaches, particularly in the SUMO environment. As shown in Figure \ref{fig:synthetic_envs}, MAA2C and QMIX exhibit higher rewards over time, indicating their ability to effectively manage traffic in complex scenarios where dynamic routing and multiple traffic flows are present. SUMO’s dynamic routing features allow centralized algorithms to optimize traffic signal timings more effectively, leading to reduced delays and shorter queue lengths.

Conversely, in CityFlow, where vehicle dynamics are simpler and there is no dynamic routing, the performance gap between centralized and decentralized methods narrows. All algorithms perform similarly, reflecting the reduced need for complex coordination in this simulation environment. This suggests that CityFlow, while computationally efficient, may not capture the same level of traffic dynamics that allow centralized algorithms to excel.

\begin{table}[h!]
    \small
    \centering
    \caption{Mean and Standard Error of Metrics for Various Controllers Across Scenarios}
    \begin{tabularx}{\textwidth}{lcccccccccc}
        \hline && \\[-0.35cm]
        \multirow{2}{*}{\textbf{Metric}} & \multicolumn{5}{c}{\textbf{MARL Controllers}} & \multicolumn{4}{c}{\textbf{Rule-based Controllers}} \\
        & \textbf{IQL} & \textbf{IA2C} & \textbf{VDN} & \textbf{QMIX} & \textbf{MAA2C} & \textbf{Fixed} & \textbf{Greedy} & \textbf{Max Pres.} & \textbf{SOTL}  \\ && \\[-0.35cm]
        \hline && \\[-0.35cm]
        \multicolumn{10}{c}{\textbf{$2\times2$ SUMO}} \\ && \\[-0.35cm]
        \hline && \\[-0.35cm]
        \textbf{Queue} & 486.47 & 588.11 & 475.32 & 276.69 & 333.48 & 464.87 & 556.88 & 556.88 & 556.88 \\
        \textbf{Delay} & 0.583 & 0.586 & 0.586 & 0.579 & 0.564 & 0.59 & 0.67 & 0.67 & 0.67 \\
        \textbf{Travel Time} & 280.34 & 329.45 & 274.92 & 188.78 & 239.00 & 325.02 & 261.45 & 261.45 & 261.45 \\
        \hline && \\[-0.35cm]
        \multicolumn{10}{c}{\textbf{$3\times3$ SUMO}} \\ && \\[-0.35cm]
        \hline && \\[-0.35cm]
        \textbf{Queue} & 500.98 & 411.47 & 520.54 & 361.57 & 449.05 & 560.08 & 656.85 & 656.85 & 656.85 \\
        \textbf{Delay} & 0.585 & 0.556 & 0.588 & 0.534 & 0.554 & 0.61 & 0.62 & 0.62 & 0.62 \\
        \textbf{Travel Time} & 140.31 & 139.92 & 137.19 & 146.76 & 143.27 & 180.52 & 166.68 & 166.68 & 166.68 \\
        \hline && \\[-0.35cm]
        \multicolumn{10}{c}{\textbf{$2\times2$ CityFlow}} \\ && \\[-0.35cm]
        \hline && \\[-0.35cm]
        \textbf{Queue} & 222.72 & 201.71 & 239.34 & 224.97 & 205.05 & 314.08 & 281.75 & 261.41 & 377.43 \\
        \textbf{Delay} & 0.606 & 0.600 & 0.611 & 0.605 & 0.598 & 0.6603 & 0.6466 & 0.6362 & 0.6993 \\
        \textbf{Travel Time} & 223.13 & 191.88 & 239.19 & 227.97 & 209.88 & 325.02 & 219.48 & 203.31 & 437.33 \\
        \hline && \\[-0.35cm]
        \multicolumn{10}{c}{\textbf{$3\times3$ CityFlow}} \\ && \\[-0.35cm]
        \hline && \\[-0.35cm]
        \textbf{Queue} & 445.10 & 396.93 & 520.17 & 484.57 & 404.01 & 621.36 & 557.96 & 458.11 & 738.95 \\
        \textbf{Delay} & 0.65 & 0.63 & 0.67 & 0.68 & 0.64 & 0.69 & 0.66 & 0.64 & 0.74 \\
        \textbf{Travel Time} & 285.17 & 249.42 & 312.03 & 290.67 & 288.64 & 389.21 & 277.8 & 246.52 & 498.33 \\
        \hline && \\[-0.35cm]
    \end{tabularx}
    \label{tab:multirow_scenario_results}
\end{table}

\subsection{Impact of Real-World Topology on Algorithm Performance}

The evaluation on real-world networks reveals important insights into how network topology influences algorithm performance.

\begin{figure}[!ht]
    \centering
    \begin{subfigure}{0.375\textwidth}
        \includegraphics[width=\textwidth]{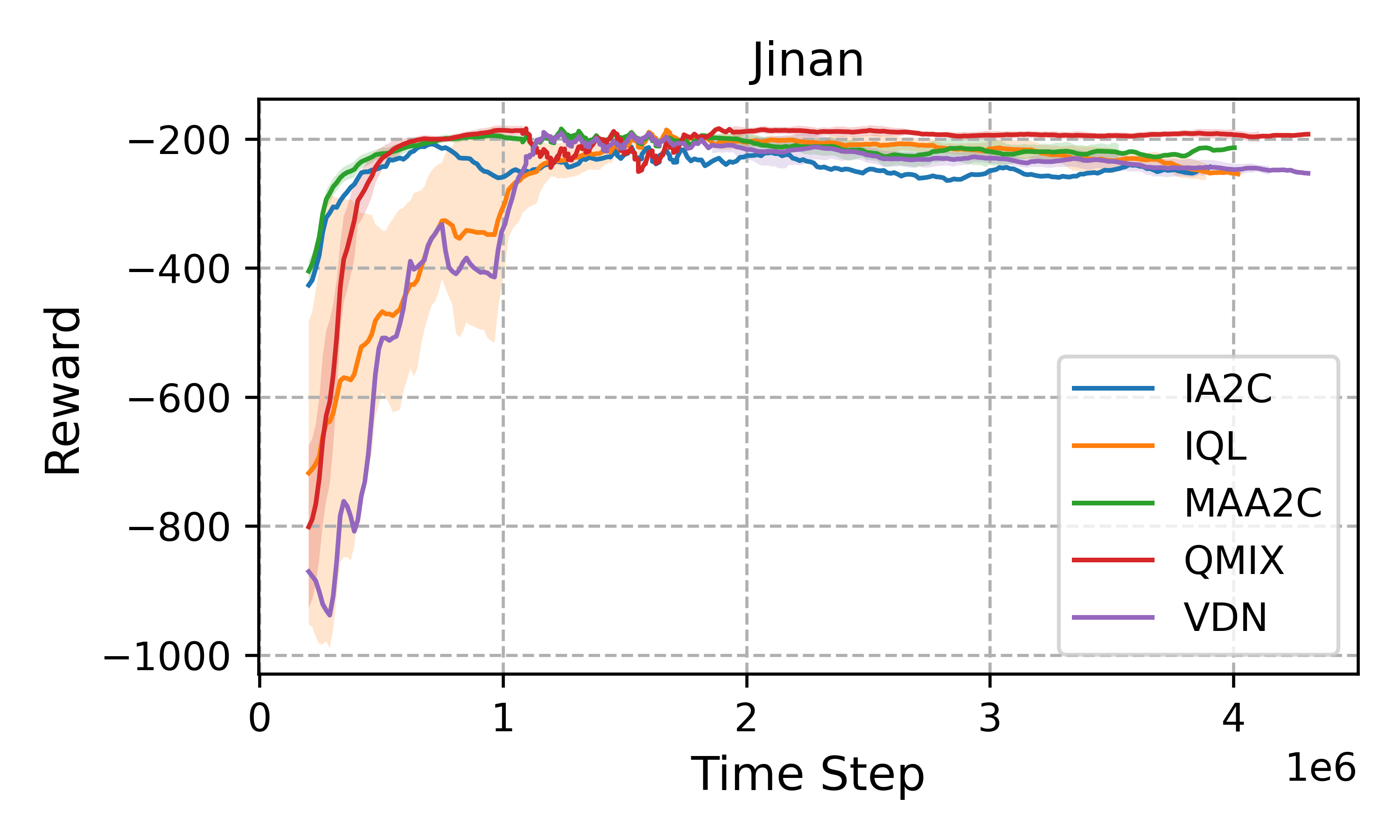}
        \label{fig:jinan_rewards}
    \end{subfigure}
    \begin{subfigure}{0.375\textwidth}
        \includegraphics[width=\textwidth]{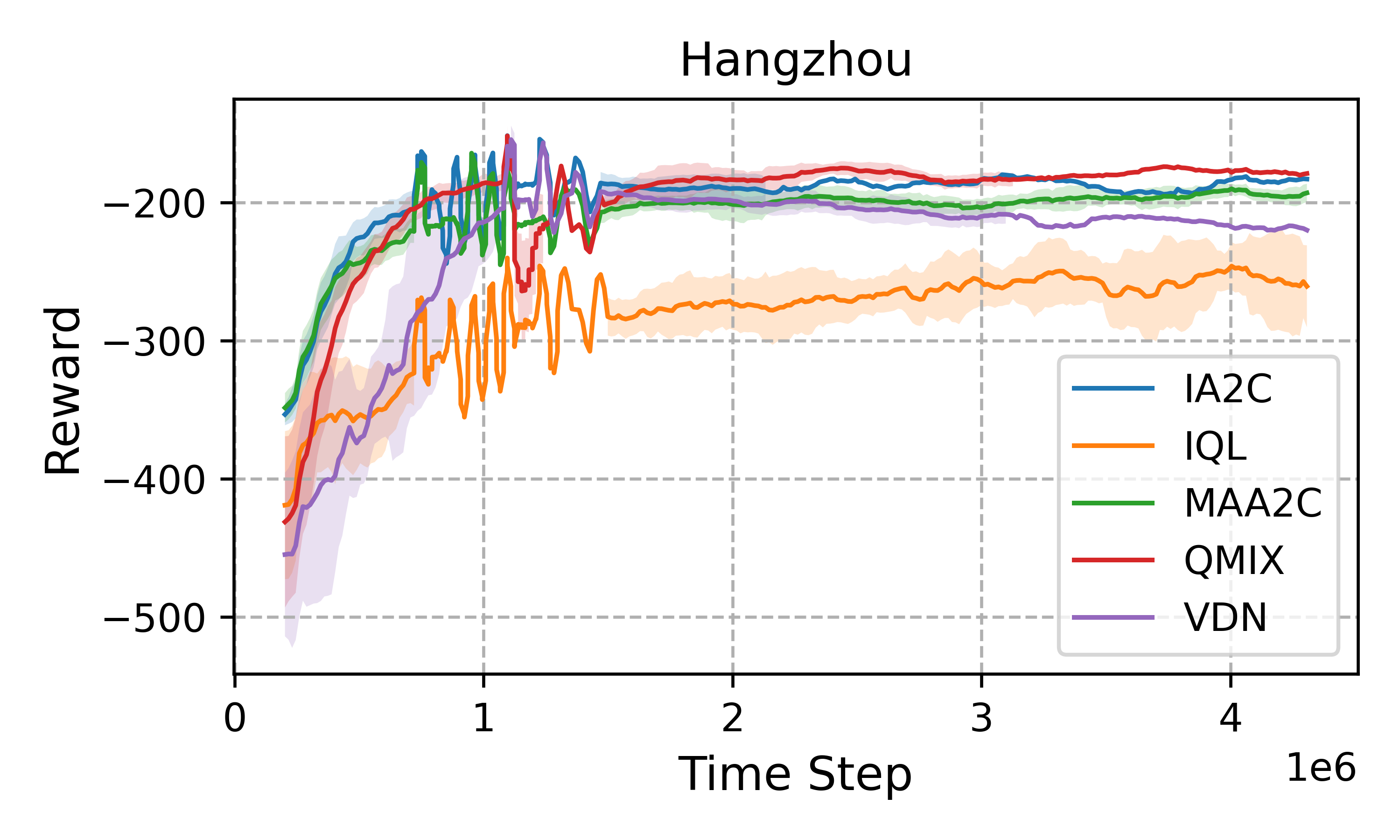}
        \label{fig:hangzhou_rewards}
    \end{subfigure}
\end{figure}

\paragraph{Jinan and Hangzhou:} In these larger, more congested networks, centralized algorithms like QMIX and MAA2C outperform decentralized approaches such as IQL and IA2C. The results highlight the importance of centralized training across traffic signals is critical for managing complex flows of vehicles in dense urban environments. The superior performance of centralized methods in these scenarios reflects their ability to adapt traffic signals in a coordinated manner, optimizing routes and reducing overall delay.

\begin{table}[h!]
    \small
    \centering
    \caption{Mean and Standard Error of Metrics for Various Controllers Across Scenarios}
    \begin{tabularx}{0.98\textwidth}{lcccccccccc}
        \hline && \\[-0.35cm]
        \multirow{2}{*}{\textbf{Metric}} & \multicolumn{5}{c}{\textbf{MARL Controllers}} & \multicolumn{4}{c}{\textbf{Rule-based Controllers}} \\
        & \textbf{IQL} & \textbf{IA2C} & \textbf{VDN} & \textbf{QMIX} & \textbf{MAA2C} & \textbf{Fixed} & \textbf{Greedy} & \textbf{Max Pres.} & \textbf{SOTL}  \\ && \\[-0.35cm]
        \hline && \\[-0.35cm]
        \multicolumn{10}{c}{\textbf{Jinan}} \\ && \\[-0.35cm]
        \hline && \\[-0.35cm]
        \textbf{Queue} & 278.55 & 248.49 & 269.14 & 231.56 & 221.79 & 413.55 & 210.17 & 228.2 & 663.39 \\
        \textbf{Delay} & 0.479 & 0.467 & 0.476 & 0.469 & 0.470 & 0.54 & 0.47 & 0.49 & 0.53 \\
        \textbf{Travel Time} & 320.51 & 311.91 & 323.85 & 308.67 & 307.00 & 354.96 & 287.33 & 294.97 & 425.01 \\
        \hline && \\[-0.35cm]
        \multicolumn{10}{c}{\textbf{Hangzhou}} \\ && \\[-0.35cm]
        \hline && \\[-0.35cm]
        \textbf{Queue} & 286.11 & 203.24 & 234.48 & 208.95 & 211.18 & 301.93 & 171.16 & 177.93 & 354.57 \\
        \textbf{Delay} & 0.585 & 0.548 & 0.565 & 0.554 & 0.559 & 0.63 & 0.56 & 0.58 & 0.61 \\
        \textbf{Travel Time} & 344.72 & 315.13 & 327.71 & 319.03 & 319.11 & 356.65 & 292.77 & 296.73 & 364.87 \\
        \hline && \\[-0.35cm]
    \end{tabularx}
    \label{tab:multirow_scenario_results}
\end{table}

\begin{figure}[!ht]
    \centering
    \begin{subfigure}{0.375\textwidth}
        \includegraphics[width=\textwidth]{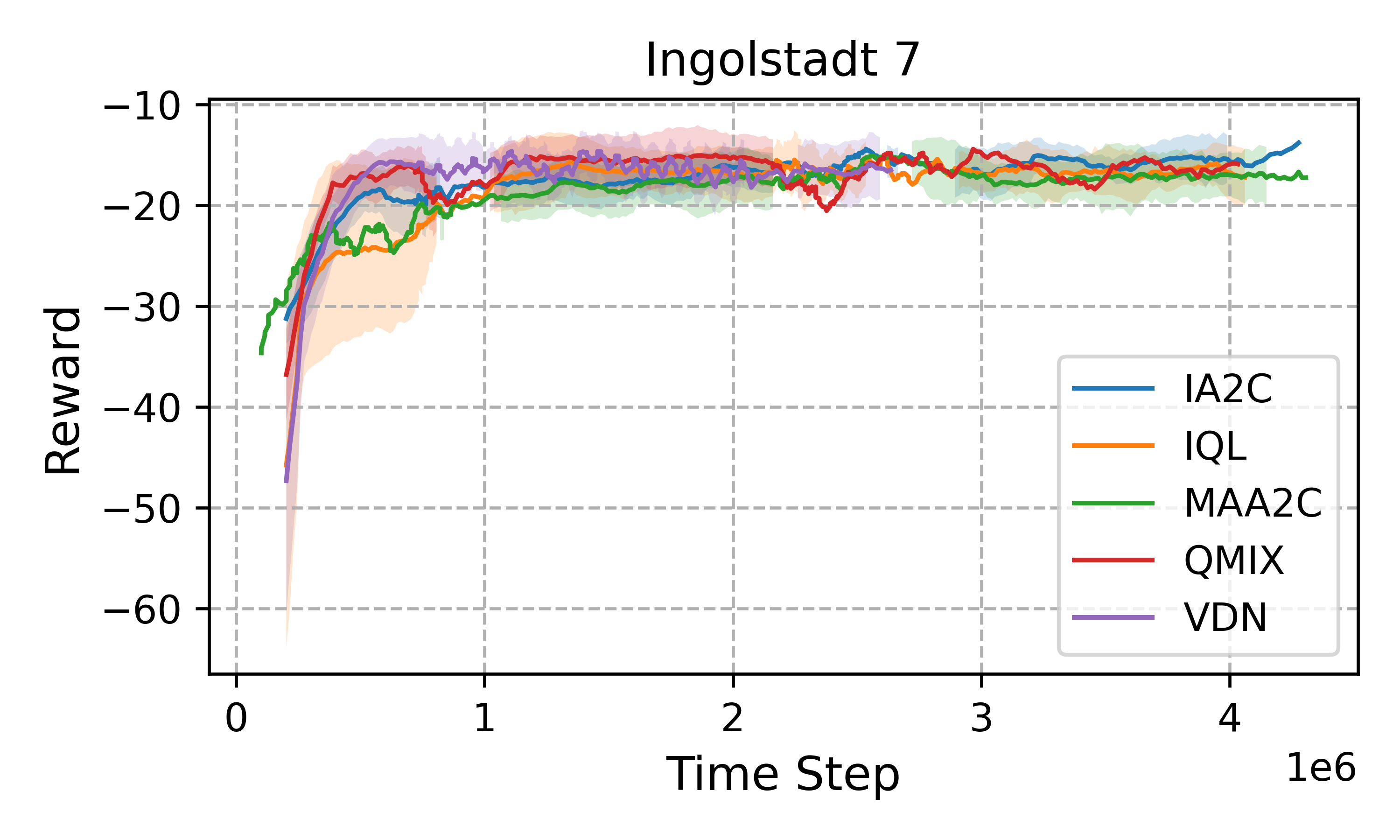}
        \label{fig:jinan_rewards}
    \end{subfigure}
    \begin{subfigure}{0.375\textwidth}
        \includegraphics[width=\textwidth]{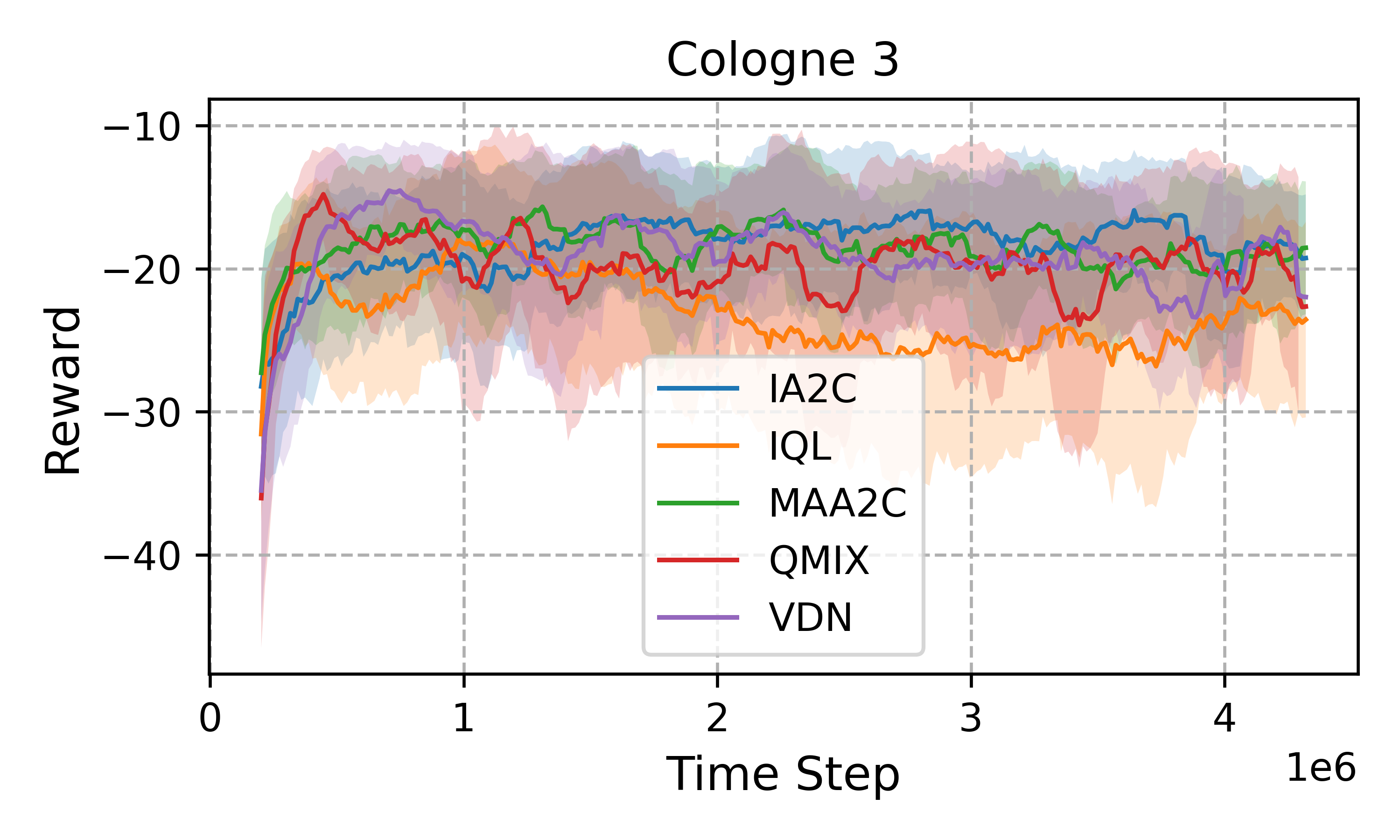}
        \label{fig:hangzhou_rewards}
    \end{subfigure}
\end{figure}

\paragraph{Cologne (3 and 8 Traffic Signals):} The performance in the Cologne networks shows the influence of topology on algorithm effectiveness. In the simpler, 3-signal Cologne network (Figure \ref{fig:cologne_network}), decentralized algorithms perform competitively with centralized methods. The linear structure and reduced complexity of the network limit the need for extensive coordination, allowing decentralized approaches to function well. However, in the more complex 8-signal Cologne network (Figure \ref{fig:cologne8_network}), centralized methods like QMIX and MAA2C show clear advantages, managing traffic across the wider grid more effectively and leading to higher rewards and lower delays.

\paragraph{Ingolstadt and Pasubio:} In these irregular and sprawling networks, centralized methods outperform decentralized ones. The elongated structure of Ingolstadt (Figure \ref{fig:ingolstadt_network}) and the spread-out intersections in Pasubio (Figure \ref{fig:pasubio_network}) introduce challenges for traffic management, such as bottlenecks and uneven traffic distribution. Centralized algorithms, with their ability to coordinate across signals and optimize traffic flow holistically, significantly reduce queues and delays in these environments. In contrast, decentralized methods struggle to maintain consistent performance in such irregular topologies, resulting in higher travel times and delays.

\begin{table}[h!]
    \small
    \centering
    \caption{Mean and Standard Error of Metrics for Various Controllers Across Scenarios}
    \begin{tabularx}{0.98\textwidth}{lcccccccccc}
        \hline && \\[-0.35cm]
        \multirow{2}{*}{\textbf{Metric}} & \multicolumn{5}{c}{\textbf{MARL Controllers}} & \multicolumn{4}{c}{\textbf{Rule-based Controllers}} \\
        & \textbf{IQL} & \textbf{IA2C} & \textbf{VDN} & \textbf{QMIX} & \textbf{MAA2C} & \textbf{Fixed} & \textbf{Greedy} & \textbf{Max Pres.} & \textbf{SOTL}  \\
        \hline && \\[-0.35cm]
        \multicolumn{10}{c}{\textbf{Cologne 3}} \\ && \\[-0.35cm]
        \hline && \\[-0.35cm]
        \textbf{Queue} & 23.46 & 18.67 & 19.55 & 20.49 & 18.80 & 51.52 & 53.29 & 53.29 & 53.29 \\
        \textbf{Delay} & 0.334 & 0.300 & 0.316 & 0.319 & 0.301 & 0.5 & 0.54 & 0.54 & 0.54 \\
        \textbf{Travel Time} & 227.11 & 227.23 & 230.06 & 227.46 & 230.87 & 220.7 & 210.54 & 210.54 & 210.54 \\
        \hline && \\[-0.35cm]
        \multicolumn{10}{c}{\textbf{Cologne 8}} \\ && \\[-0.35cm]
        \hline && \\[-0.35cm]
        \textbf{Queue} & 19.95 & 18.32 & 20.77 & 17.35 & 17.39 & 41.71 & 73.87 & 73.87 & 73.87 \\
        \textbf{Delay} & 0.190 & 0.185 & 0.184 & 0.171 & 0.183 & 0.26 & 0.3 & 0.3 & 0.3 \\
        \textbf{Travel Time} & 356.59 & 356.23 & 356.08 & 356.96 & 356.52 & 358.14 & 339.59 & 339.59 & 339.59 \\
        \hline && \\[-0.35cm]
        \multicolumn{10}{c}{\textbf{Ingolstadt 7}} \\ && \\[-0.35cm]
        \hline && \\[-0.35cm]
        \textbf{Queue} & 18.45 & 17.46 & 20.35 & 17.38 & 19.19 & 107.06 & 51.8 & 50.56 & 44.16 \\
        \textbf{Delay} & 0.184 & 0.169 & 0.190 & 0.173 & 0.177 & 0.42 & 0.37 & 0.36 & 0.33 \\
        \textbf{Travel Time} & 192.82 & 193.35 & 193.04 & 193.03 & 192.74 & 174.66 & 183.37 & 184.47 & 186.09 \\
        \hline && \\[-0.35cm]
        \multicolumn{10}{c}{\textbf{Pasubio}} \\ && \\[-0.35cm]
        \hline && \\[-0.35cm]
        \textbf{Queue} & 335.29 & 340.11 & 307.60 & 297.79 & 322.94 & 304.82 & 330.26 & 330.26 & 330.26 \\
        \textbf{Delay} & 0.35 & 0.33 & 0.30 & 0.29 & 0.30 & 0.34 & 0.39 & 0.39 & 0.39 \\
        \textbf{Travel Time} & 303.67 & 310.88 & 322.29 & 321.03 & 327.31 & 326.0 & 292.09 & 292.09 & 292.09 \\
        \hline && \\[-0.35cm]
    \end{tabularx}
    \label{tab:multirow_scenario_results}
\end{table}

\subsection{Insights from Performance Metrics}

Table \ref{tab:multirow_scenario_results} summarizes key performance metrics, such as queue lengths, delays, and travel times, across the evaluated networks. The results reveal several key trends:

\begin{itemize}
    \item \textbf{Queue Length:} Centralized algorithms generally achieve shorter queue lengths, particularly in more complex networks like Pasubio and Cologne 8, where coordination across intersections is critical for maintaining traffic flow.
    
    \item \textbf{Delays:} Across all scenarios, centralized approaches reduce delays more effectively than decentralized ones. The difference is especially notable in SUMO environments, where dynamic routing allows centralized algorithms to adapt to real-time traffic conditions more effectively.
    
    \item \textbf{Travel Time:} In networks with higher complexity, such as Jinan and Hangzhou, centralized approaches minimize travel time more effectively than decentralized methods. However, in simpler networks like CityFlow’s 2x2 grid, the travel times across all algorithms are comparable, suggesting that centralized coordination offers diminishing returns in less complex environments.
\end{itemize}

\subsection{Topology-Dependent Performance}

The topology of the traffic network plays a crucial role in determining the success of different MARL algorithms. In simpler, grid-like networks, such as the 3-signal Cologne setup, decentralized algorithms can perform on par with centralized methods. However, as network complexity increases, the advantages of centralized coordination become more pronounced. In networks like Pasubio and Ingolstadt, which feature irregular layouts and complex traffic flows, centralized approaches like QMIX and MAA2C outperform decentralized methods by a significant margin.

This suggests that the choice of algorithm should be informed by the specific characteristics of the traffic network. Centralized algorithms are more suitable for complex, irregular networks with high traffic density, while decentralized approaches may suffice in simpler, more uniform networks.

\section{Conclusion}

In this work, we introduced PyTSC, a versatile and extensible library designed to address the significant gaps in MARL-based Traffic Signal Control (TSC) research. By enabling compatibility with multiple simulators, such as SUMO and CityFlow, PyTSC provides a unified API that simplifies the development, testing, and benchmarking of Multi-Agent Reinforcement Learning (MARL) algorithms for TSC. Additionally, its optimized architecture facilitates faster simulation speeds, allowing researchers to focus on algorithmic innovations rather than technical integration.

PyTSC's integration with modern CTDE (Centralized Training Decentralized Execution) frameworks, such as EPyMARL and MARLLib, allows researchers to prototype and evaluate advanced MARL techniques within a traffic control setting. This fills a critical gap in the existing research ecosystem, where tools are either too domain-specific or lack the flexibility required for seamless experimentation.

With its modular and extensible design, PyTSC establishes a foundation for more consistent, reproducible, and scalable MARL research in the realm of traffic signal control. This contributes directly to advancing smarter, more adaptive traffic management solutions. The performance evaluations across both synthetic and real-world traffic networks demonstrate the library’s effectiveness and versatility, underscoring its potential to drive innovation in TSC systems.

\section{Future Work}

Looking ahead, PyTSC opens several avenues for future development and exploration:
\begin{itemize}
    \item \textbf{Incorporation of Additional MARL Algorithms:} Future iterations of PyTSC will include policy-based MARL algorithms such as MADDPG, IPPO, and MAPPO, expanding its capabilities for more complex traffic control strategies.
    \item \textbf{Diverse Traffic Flow Generation:}  We aim to introduce more diverse traffic flow generation models, including both synthetic and real-world flow patterns, to better simulate the variety of traffic scenarios seen in urban environments.
    \item \textbf{Synthetic Network Generation Modules:} Expanding the tools for generating synthetic traffic signal networks will enable researchers to test MARL algorithms in even more controlled and complex environments, beyond the existing benchmarks.
    \item \textbf{Tailored Neural Network Architectures:} While the current version of PyTSC relies on out-of-the-box neural architectures, future updates will explore the integration of specialized models, such as Graph Neural Networks (GNNs), that can leverage the graphical structure of traffic networks to optimize decision-making.
    \item \textbf{Extending Simulator Compatibility:} We also plan to integrate additional traffic simulators into the PyTSC framework, ensuring that the platform remains adaptable to emerging technologies and research needs in the field of intelligent transportation systems.
\end{itemize}
By continuously evolving PyTSC, we aim to provide a robust and dynamic platform that will foster innovation in MARL-based traffic signal control and further enhance the real-world applicability of these technologies.
\newpage

\bibliography{references}

\begin{thebibliography}{10}

\bibitem{du2021survey}
W.~Du and S.~Ding, ``A survey on multi-agent deep reinforcement learning: from
  the perspective of challenges and applications,'' {\em Artificial
  Intelligence Review}, vol.~54, pp.~3215--3238, 2021.

\bibitem{nguyen2020deep}
T.~T. Nguyen, N.~D. Nguyen, and S.~Nahavandi, ``Deep reinforcement learning for
  multiagent systems: A review of challenges, solutions, and applications,''
  {\em IEEE transactions on cybernetics}, vol.~50, no.~9, pp.~3826--3839, 2020.

\bibitem{canese2021multi}
L.~Canese, G.~C. Cardarilli, L.~Di~Nunzio, R.~Fazzolari, D.~Giardino, M.~Re,
  and S.~Span{\`o}, ``Multi-agent reinforcement learning: A review of
  challenges and applications,'' {\em Applied Sciences}, vol.~11, no.~11,
  p.~4948, 2021.

\bibitem{noaeen2022reinforcement}
M.~Noaeen, A.~Naik, L.~Goodman, J.~Crebo, T.~Abrar, Z.~S.~H. Abad, A.~L.
  Bazzan, and B.~Far, ``Reinforcement learning in urban network traffic signal
  control: A systematic literature review,'' {\em Expert Systems with
  Applications}, vol.~199, p.~116830, 2022.

\bibitem{chen2022real}
R.~Chen, F.~Fang, and N.~Sadeh, ``The real deal: A review of challenges and
  opportunities in moving reinforcement learning-based traffic signal control
  systems towards reality,'' {\em arXiv preprint arXiv:2206.11996}, 2022.

\bibitem{wei2019survey}
H.~Wei, G.~Zheng, V.~Gayah, and Z.~Li, ``A survey on traffic signal control
  methods,'' {\em arXiv preprint arXiv:1904.08117}, 2019.

\bibitem{haydari2020deep}
A.~Haydari and Y.~Y{\i}lmaz, ``Deep reinforcement learning for intelligent
  transportation systems: A survey,'' {\em IEEE Transactions on Intelligent
  Transportation Systems}, vol.~23, no.~1, pp.~11--32, 2020.

\bibitem{SUMO2018}
P.~A. Lopez, M.~Behrisch, L.~Bieker-Walz, J.~Erdmann, Y.-P. Fl{\"o}tter{\"o}d,
  R.~Hilbrich, L.~L{\"u}cken, J.~Rummel, P.~Wagner, and E.~Wie{\ss}ner,
  ``Microscopic traffic simulation using sumo,'' in {\em The 21st IEEE
  International Conference on Intelligent Transportation Systems}, IEEE, 2018.

\bibitem{zhang2019cityflow}
H.~Zhang, S.~Feng, C.~Liu, Y.~Ding, Y.~Zhu, Z.~Zhou, W.~Zhang, Y.~Yu, H.~Jin,
  and Z.~Li, ``Cityflow: A multi-agent reinforcement learning environment for
  large scale city traffic scenario,'' in {\em The world wide web conference},
  pp.~3620--3624, 2019.

\bibitem{sumorl}
L.~N. Alegre, ``{SUMO-RL}.'' \url{https://github.com/LucasAlegre/sumo-rl},
  2019.

\bibitem{ault2021reinforcement}
J.~Ault and G.~Sharon, ``Reinforcement learning benchmarks for traffic signal
  control,'' in {\em Thirty-fifth Conference on Neural Information Processing
  Systems Datasets and Benchmarks Track (Round 1)}, 2021.

\bibitem{mei2022libsignal}
H.~Mei, X.~Lei, L.~Da, B.~Shi, and H.~Wei, ``Libsignal: An open library for
  traffic signal control,'' {\em arXiv preprint arXiv:2211.10649}, 2022.

\bibitem{papoudakis2020benchmarking}
G.~Papoudakis, F.~Christianos, L.~Sch{\"a}fer, and S.~V. Albrecht,
  ``Benchmarking multi-agent deep reinforcement learning algorithms in
  cooperative tasks,'' {\em arXiv preprint arXiv:2006.07869}, 2020.

\bibitem{hu2022marllib}
S.~Hu, Y.~Zhong, M.~Gao, W.~Wang, H.~Dong, Z.~Li, X.~Liang, Y.~Yang, and
  X.~Chang, ``Marllib: Extending rllib for multi-agent reinforcement
  learning,'' 2022.

\bibitem{liang2018rllib}
E.~Liang, R.~Liaw, R.~Nishihara, P.~Moritz, R.~Fox, K.~Goldberg, J.~E.
  Gonzalez, M.~I. Jordan, and I.~Stoica, ``{RLlib}: Abstractions for
  distributed reinforcement learning,'' in {\em International Conference on
  Machine Learning ({ICML})}, 2018.

\bibitem{samvelyan2019starcraft}
M.~Samvelyan, T.~Rashid, C.~S. De~Witt, G.~Farquhar, N.~Nardelli, T.~G. Rudner,
  C.-M. Hung, P.~H. Torr, J.~Foerster, and S.~Whiteson, ``The starcraft
  multi-agent challenge,'' {\em arXiv preprint arXiv:1902.04043}, 2019.

\bibitem{oliehoek2016concise}
F.~A. Oliehoek and C.~Amato, {\em A concise introduction to decentralized
  POMDPs}.
\newblock Springer International Publishing, Cham, 2016.

\end{thebibliography}
\bibliographystyle{ieeetr}

\end{document}